# A simple stress-dilatancy equation for sand

**Yifei Sun**

Associate Professor, College of Civil and Transportation Engineering, Hohai University, China;
Humboldt Fellow, Faculty of Civil and Environmental Engineering, Ruhr Universität Bochum, Germany
E-mail: yifei.sun@rub.de

## ABSTRACT

The stress-dilatancy relation is of critical importance for constitutive modelling of sand. A new fractional-order stress-dilatancy equation is analytically developed in this study, based on stress-fractional operators. An apparent linear response of the stress-dilatancy behaviour of soil after sufficient shearing is obtained. As the fractional order varies, the derived stress-dilatancy curve and the associated phase transformation state stress ratio shift. But, unlike existing researches, no other specific parameters, except the fractional order, concerning such shift and the state-dependence are required. The developed stress-dilatancy equation is then incorporated into an existing constitutive model for validation. Test results of different sands are simulated and compared, where a good model performance is observed.

# 1. Introduction

The constitutive model for geomaterial is an essential element for numerical analysis of the strength and deformation characteristics of geotechnical facilities. In practical finite element modelling, the rate-independent stress-strain behavior of geomaterial was often simulated by using an elastoplastic approach (Sumelka and Nowak, 2016; Shi et al., 2019; 2020). It has been recognized that the plastic flow of soil, e.g., sand (Wichtmann and Triantafyllidis, 2016) and clay (Tafili and Triantafyllidis, 2020), was dependent on its material state. To capture such state-dependent behavior of geomaterials, different stress-dilatancy equations incorporating various state parameters were proposed, for example, the state-dependent Rowe's stress-dilatancy equation (Wan and Guo, 1998) using '$e/e_c$', where $e$ and $e_c$ are void ratios at the current and critical states, respectively. In addition, Sumelka and Nowak (2016) developed a fractional plastic flow rule for capturing the induced plastic anisotropy of civil materials. Sun et al. (2019b) explored one mathematical principle underlying the state-dependent stress-dilatancy behavior of geomaterial, by using stress-fractional operators. Despite of the good descriptions of the state dependent stress-strain behavior of geomaterial, their approach was found to have negative performance when simulating the stress-dilatancy data of sand.

Therefore, in this study, an attempt is made to provide a modified fractional-order stress-dilatancy equation, based on the original Cam-clay yielding function proposed in Schofield and Wroth (1968). The triaxial test results of different sands will be simulated and compared by using the proposed fractional-order stress-dilatancy equation. The study is structured as follows: Section 2 develops a fractional-order stress-dilatancy equation, where detailed analytical solutions are provided; model application is shown in Section 3; Section 4 concludes the study.

# 2. Stress-dilatancy Equation

In this study, the Cam-clay yielding function ($f$) (Schofield and Wroth, 1968) is used.

$$f = q + Mp'\ln p' - Mp'\ln p'_0 = 0 \quad (1)$$

where $p'_0$ is the intercept of $f$ with the $p'$-axis. In triaxial loading condition, the mean effective principal stress $p' = (\sigma'_1 + 2\sigma'_3)/3$, and the generalised shear stress $q = \sigma'_1 - \sigma'_3$, where $\sigma'_1$ and $\sigma'_3$ are the first and third effective stresses, respectively. The stress ratio $\eta = q/p'$. Furthermore, the corresponding volumetric strain $\varepsilon_v = \varepsilon_1 + 2\varepsilon_3$ and the generalised shear strain $\varepsilon_s = $

$2/3(\varepsilon_1 - \varepsilon_3)$, where $\varepsilon_1$ and $\varepsilon_3$ are the first and third principal strains, respectively. $M$ is the critical-state stress ratio in the $p' - q$ plane, which can be defined as: $M = 6\sin\varphi_c/(3t - \sin\varphi_c)$, where $t = +1$ or $-1$ for compression or extension, respectively; $\varphi_c$ is the critical-state friction angle under triaxial compression. Substituting Eq. (1) into Eqs. (A3) – (A4), one can obtain the expression for $d_g$:

$$d_g = \frac{\partial^\alpha f/\partial p'^\alpha}{\partial^\alpha f/\partial q^\alpha} \tag{2}$$

Following Sun et al. (2019a), the denominator can be provided as: $\partial^\alpha f/\partial q^\alpha = t|q|^{1-\alpha}/\Gamma(2-\alpha)$, where $\Gamma$ is the gamma function, defined in Eq. (A1). Note that for compressive loading, the left-sided fractional derivative defined in Eq. (A3) is used, while the right-sided one defined in Eq. (A4) is used for extensive loading. However, whatever the loading is, a unified analytical solution will be obtained. As for sand, $p' \geq 0$, throughout the test, $\partial^\alpha f/\partial p'^\alpha$ can be solved as:

$$\begin{aligned}\frac{\partial^\alpha f}{\partial p'^\alpha} &= tM\frac{\partial^\alpha(p'\ln p')}{\partial p'^\alpha} - tM\ln p'_0\frac{\partial^\alpha p'}{\partial p'^\alpha}\\ &= \frac{tp'^{1-\alpha}}{\Gamma(2-\alpha)}\{[\psi(2) - \psi(2-\alpha)]M - \eta\}\end{aligned} \tag{3}$$

where $\psi$ is the digamma function, defined in Eq. (A2). Then, Eq. (2) can be further derived as

$$d_g = (\mu M - \eta)|\eta|^{1-\alpha} \tag{4}$$

where the coefficient $\mu = \psi(2) - \psi(2 - \alpha)$. It is found that when $\alpha = 1$, $\mu = 1$ and Eq. (4) reduces to the original Cam-clay stress-dilatancy equation, i.e., $d_g = M - \eta$, which is only determined by $(M, \eta)$ and not influenced by any intermediate state or the memory from the current state to the final critical state. Thus, the predicted phase transformation state stress ratio ($\eta_d$) is always the same as $M$, which is inconsistent with experimental observations (Been and Jefferies, 1985). To reflect the effect of state dependence, a varying $M_d$ coupled with a scaling scalar $d_0$, instead of the sole $M$, were usually used, e.g., $d_g = d_0/M\,(M_d - \eta)$ (Li and Dafalias, 2000), where $M_d = M\exp(n\Psi)$, $n$ is a material constant and $\Psi\,(= e - e_c)$, is the state parameter (Been and Jefferies, 1985). Note that $\Psi$ instead of $\psi$ is used here to avoid confusion with the digamma function ($\psi$). $M_d$ would be larger than $M$, when the specimen was initially at the "wet" side of the critical state line (CSL), whereas $M_d$ would be smaller than $M$, when the specimen was initially at the "dry" side of the critical state line (CSL). However, the use of a varying $M_d$ was a phenomenological mapping of the experimental results, which may not represent the mathematical origin of the state-dependence and space-shifting of $d_g$ in the $\eta - d_g$ space. Therefore, rather to use two parameters, i.e., $M_d$ and $d_0$, the variable-order fractional derivative is used in this study, which will also decrease the number of the required model parameters by one. As $\partial^\alpha f/\partial\sigma^\alpha$ are defined based on an interval, the effect of state history can be automatically considered during the mathematical derivation. It is found that the state dependence and shifting of the developed stress-dilatancy equation can be simultaneously captured by only using one scalar, i.e., the fractional order ($\alpha$). As shown in Fig. 1, the stress-dilatancy curve shifts as $\alpha$ varies. A higher phase transformation state $\eta_d$ is reported as $\alpha$ increases. According to Eq. (4), $\eta_d$ can be obtained at $d_g = 0$, such that

$$\eta_d = \mu M \qquad (5)$$

where it can be found that the coefficient $\mu$ controls $\eta_d$. Due to the non-integer feature of $\alpha$, $\eta_d \neq M$, always exists, which agrees with the laboratory observations (Wichtmann and Triantafyllidis, 2016). It is thus mathematically proved that due to the memory of the loading history, $\eta_d$ is no longer equal to the final critical-state stress ratio. $\eta_d$ was found to be varying with the material state (Been and Jefferies, 1985). To capture such state dependence, the following variable fractional order is suggested (Sun et al., 2019a): $\alpha = \exp(\Delta\Psi)$, where $\Delta$ is a material constant; Fig. 2 shows the simulations of the test results (discrete points) of Nerlerk270/1 sand (Jefferies and Been, 2015) and Karlsruhe fine sand (Wichtmann and Triantafyllidis, 2016) with different initial void ratios ($e_0$) via Eq. (5). It can be found that the proposed approach can well reproduce the dependence of the phase transformation state stress ratio on material state of different sands.

## 3. Application

The proposed stress-dilatancy equation is incorporated into the constitutive model in Li and Dafalias (2000) by replacing its original stress-dilatancy equation. For more details of the other constitutive equations, one can refer to Li and Dafalias (2000). A series of drained and undrained test results of different sands are simulated. Details of the test material and test setup can be found in each literature (Verdugo and Ishihara, 1996; Jefferies and Been, 2015; Wichtmann and Triantafyllidis, 2016) and not repeated here. Table 1 lists the model parameters for carrying out the simulations.

Figs. 3 – 4 show the model simulations of the stress-strain behavior of Toyoura sand (Verdugo and Ishihara, 1996) tested under different initial states. It can be found that the developed approach can accurately capture the undrained and drained behaviour of Toyoura

sand, where the liquefaction, partial liquefaction, non-flow, and strain hardening/softening behaviour can be well reproduced.

Figs. 5 shows the model simulations of the stress-dilatancy behaviour of Nerlerk270/1 sand (Jefferies and Been, 2015) under different initial states. It can be observed that irrespective of the initial void ratio, the simulated stress ratio increases with the decreasing dilatancy ratio until reaching the maximum dilatancy state ($d_g = d_g^{\max}$), and then it decreases as the dilatancy ratio increases, which agrees very well the corresponding test results.

Fig. 6 compares the simulated and test results of Karlsruhe fine sand (Wichtmann and Triantafyllidis, 2016) with different initial states. It can be found that the model can well characterise the stress-dilatancy behaviour of Karlsruhe fine sand. The simulated stress ratio exhibits an initial linear increase followed by a decrease with the varying dilatancy ratio.

## 4. Conclusions

This study made an attempt to develop a new fractional-order stress-dilatancy equation for sand. The main findings can be summarized as follows:

(1) Analytical solutions of the fractional-order derivatives of the Cam-clay yielding function under compression and extension were obtained, which further led to the development of a unified fractional-order stress-dilatancy equation for sand.
(2) The derived stress-dilatancy equation was influenced by the critical-state stress ratio, current-state stress ratio and fractional order. As the fractional order increased, the stress-dilatancy curve shifted. A higher phase transformation state stress ratio was found with a higher fractional order. Model parameters can be decreased by one whilst it can still consider such space-shift of the stress-dilatancy curve.
(3) Further simulations of a series of test results of different sands showed that the developed approach can accurately reproduce the state-dependent stress-dilatancy behavior of sand.

## Acknowledgement

The financial support provided by the National Natural Science Foundation of China (Grant No. 51890912) and the Alexander Von Humboldt Foundation, Germany, are appreciated.

# Appendix

(a) Gamma function and Digamma function

$$\Gamma(x) = \int_0^\infty t^{x-1}\exp(-t)dt \tag{A1}$$

$$\psi(x) = \frac{\Gamma'(x)}{\Gamma(x)} \tag{A2}$$

where $x > 0$, is the independent variable; $\Gamma'(x)$ is the first-order derivative of the gamma function.

(b) Riemann-Liouville fractional derivative

The left-sided and right-sided Riemann-Liouville derivatives of a function $f$ are respectively defined as:

$$_{0^+}D_x^\alpha f(x) = \frac{1}{\Gamma(n-\alpha)} \frac{d^n}{dx^n} \int_{0^+}^x \frac{f(\tau)d\tau}{(x-\tau)^{\alpha+1-n}}, \quad x > 0 \tag{A3}$$

$$_xD_{0^-}^\alpha f(x) = \frac{(-1)^n}{\Gamma(n-\alpha)} \frac{d^n}{dx^n} \int_x^{0^-} \frac{f(\tau)d\tau}{(\tau-x)^{\alpha+1-n}}, \quad x < 0 \tag{A4}$$

where $D$ ($= \partial^\alpha/\partial x^\alpha$) means partial derivation; $\alpha \in (n-1, n)$, is the fractional order; $n$ is a positive integer. In addition, the CSL, i.e., $e = e_\Gamma - \lambda \ln p'$, is used for Nerlerk270/1 sand, while other sands are simulated using $e = e_\Gamma - \lambda(p'/p_a)^\xi$, where $e_\Gamma$, $\lambda$ and $\xi$ are the material constants.

# References

Been, K., and Jefferies, M. G. (1985). "A state parameter for sands." *Géotechnique*, 35(2), 99-112.
Jefferies, M., and Been, K. (2015). *Soil liquefaction: a critical state approach*, CRC press, Boca Raton, Florida, USA.
Li, X., and Dafalias, Y. (2000). "Dilatancy for cohesionless soils." *Géotechnique*, 50(4), 449-460.
Schofield, A., and Wroth, P. (1968). *Critical state soil mechanics*, McGraw-Hill London, New York, USA.
Shi, X. S., Nie, J., Zhao, J., and Gao, Y. (2020). "A homogenization equation for the small strain stiffness of gap-graded granular materials." *Comput. Geotech.*, 121, 103440.
Shi, X. S., Yin, J., and Zhao, J. (2019). "Elastic visco-plastic model for binary sand-clay mixtures with applications to one-dimensional finite strain consolidation analysis." *J. Eng. Mech.*, 145(8), 04019059.

Sumelka, W., and Nowak, M. (2016). "Non-normality and induced plastic anisotropy under fractional plastic flow rule: a numerical study." *Int. J. Numer. Anal. Meth. Geomech.*, 40(5), 651-675.

Sun, Y., Gao, Y., and Chen, C. (2019a). "Critical-state fractional model and its numerical scheme for isotropic granular soil considering state-dependence." *Int. J. Geomech.*, 13(9), 04018202.

Sun, Y., Gao, Y., and Shen, Y. (2019b). "Mathematical aspect of the state-dependent stress-dilatancy of granular soil under triaxial loading." *Géotechnique*, 69(2), 158-165.

Tafili, M., and Triantafyllidis, T. "State-dependent dilatancy of soils: experimental evidence and constitutive modeling." *Proc., Recent Developments of Soil Mechanics and Geotechnics in Theory and Practice*, Springer International Publishing, 54-84.

Verdugo, R., and Ishihara, K. (1996). "The steady state of sandy soils." *Soils Found.*, 36(2), 81-91.

Wan, R., and Guo, P. (1998). "A simple constitutive model for granular soils: modified stress-dilatancy approach." *Comput. Geotech.*, 22(2), 109-133.

Wichtmann, T., and Triantafyllidis, T. (2016). "An experimental database for the development, calibration and verification of constitutive models for sand with focus to cyclic loading: part I—tests with monotonic loading and stress cycles." *Acta Geotech.*, 11(4), 739-761.

**Table caption list**

Table 1. Model parameters

Table 1. Model parameters

| Soil type | $G_0$ | $v$ | $\varphi_c$ (°) | $\lambda$ | $e_\Gamma$ | $\xi$ | $\Delta$ | $k$ | $h_1$ ($h_2$) |
|---|---|---|---|---|---|---|---|---|---|
| Toyoura sand (Verdugo and Ishihara, 1996) | 125 | 0.05 | 31.2 | 0.019 | 0.934 | 0.7 | 1.1 | 1.1 | 3.0 |
| Nerlerk270/1 sand (Jefferies and Been, 2015) | 65 | 0.15 | 31.6 | 0.875 | 0.021 | - | 0.55 | 1.81 | 2.5 |
| Karlsruhe fine sand (Wichtmann and Triantafyllidis, 2016) | 150 | 0.05 | 33.2 | 0.122 | 1.103 | 0.205 | 0.4 | 1.42 | 2.0 |

# Figure caption list

**Fig. 1**. Effect of $\alpha$ on the stress-dilatancy curve of sand

**Fig. 2**. Observed and predicted variation of $\eta_d$ with $\Psi$ of different sands

**Fig. 3**. Model simulations of the undrained behaviour of Toyoura sand (Verdugo and Ishihara, 1996)

**Fig. 4**. Model simulations of the drained behaviour of Toyoura sand (Verdugo and Ishihara, 1996)

**Fig. 5**. Model simulations of the stress-dilatancy behaviour of Nerlerk270/1 sand (Jefferies and Been, 2015)

**Fig. 6**. Model simulations of the stress-dilatancy behaviour of loose Karlsruhe fine sand (Wichtmann and Triantafyllidis, 2016)

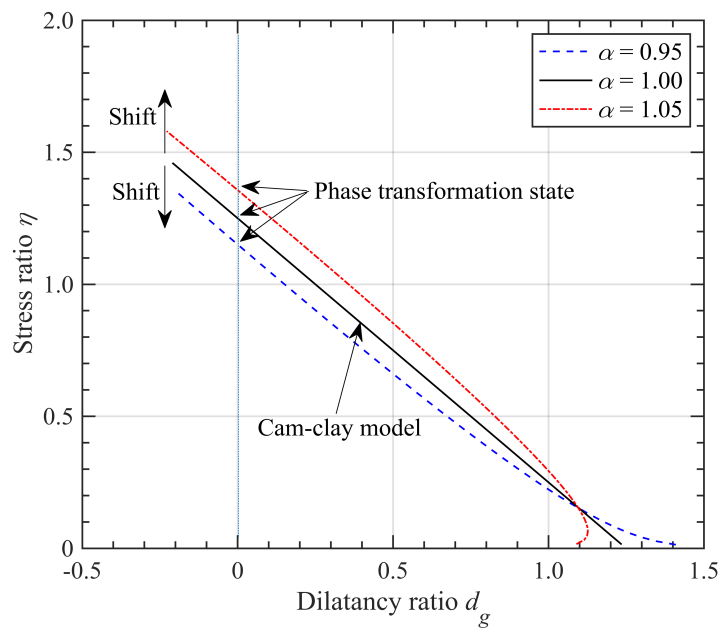

**Fig. 1**. Effect of $\alpha$ on the stress-dilatancy curve of soil

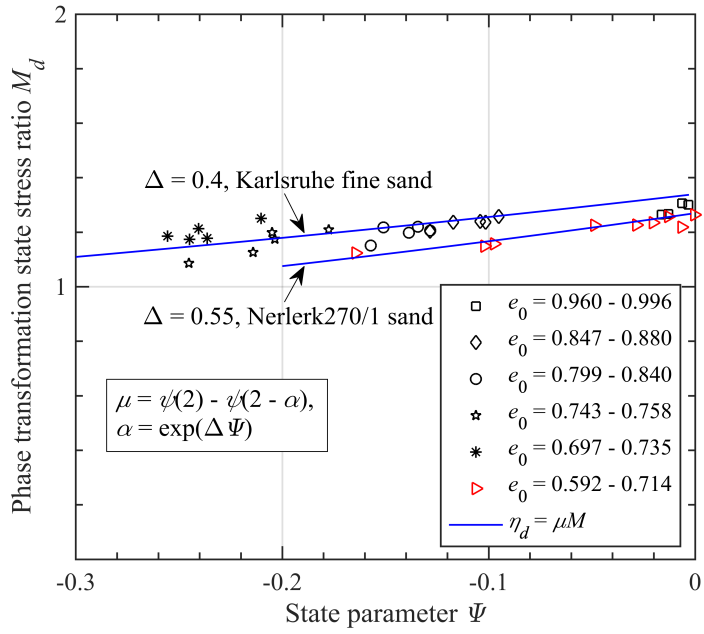

**Fig. 2**. Observed and predicted variation of $\eta_d$ with $\Psi$ of different sands

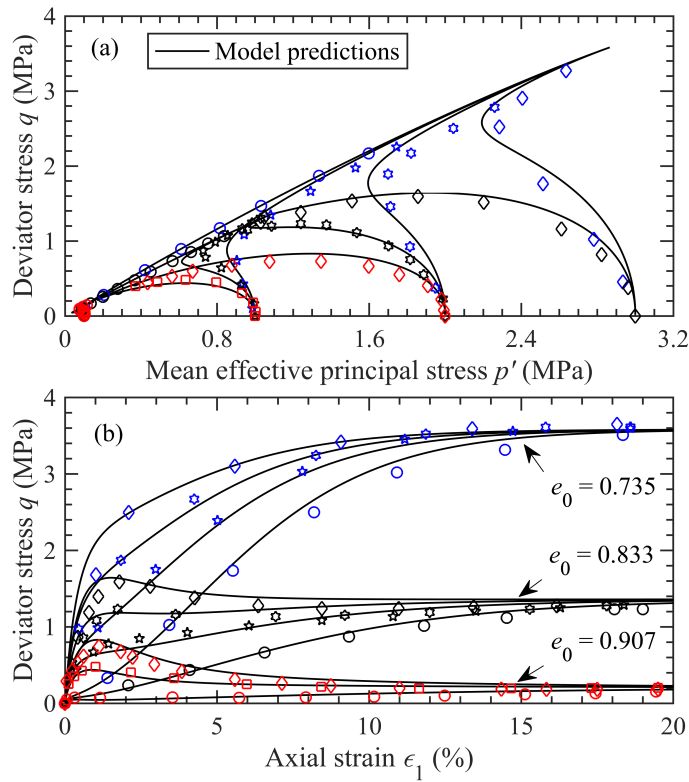

**Fig. 3**. Model simulations of the undrained behaviour of Toyoura sand (Verdugo and Ishihara, 1996)

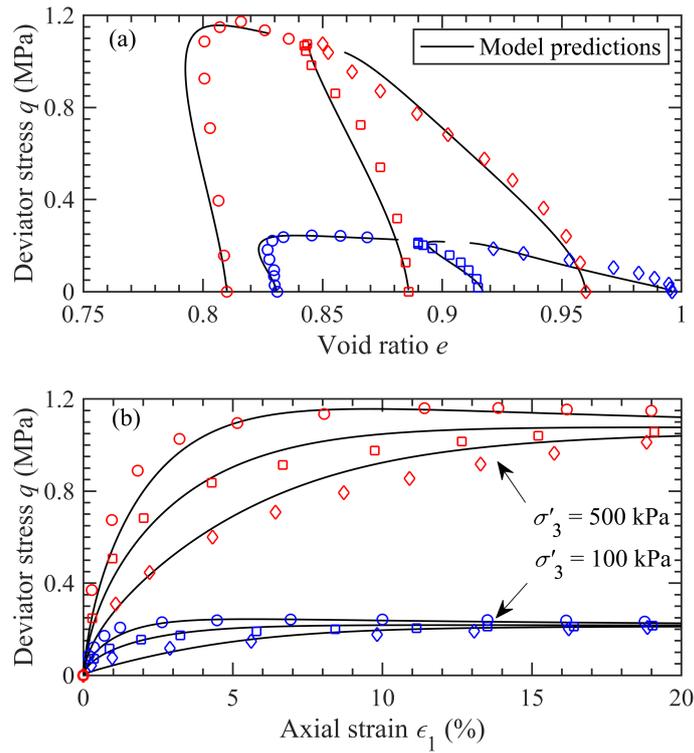

**Fig. 4**. Model simulations of the drained behaviour of Toyoura sand (Verdugo and Ishihara, 1996)

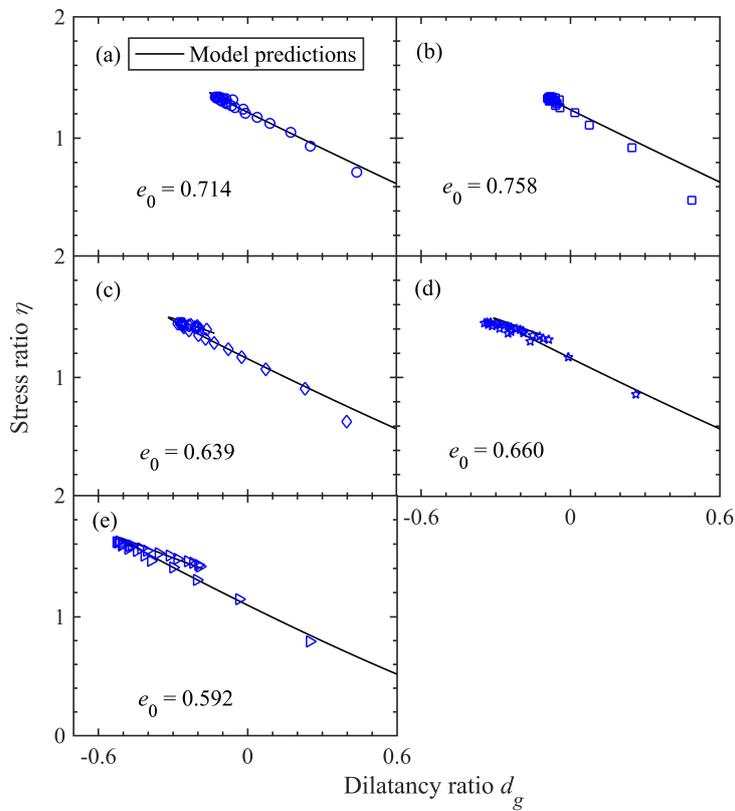

**Fig. 5**. Model simulations of the stress-dilatancy behaviour of Nerlerk270/1 sand (Jefferies and Been, 2015)

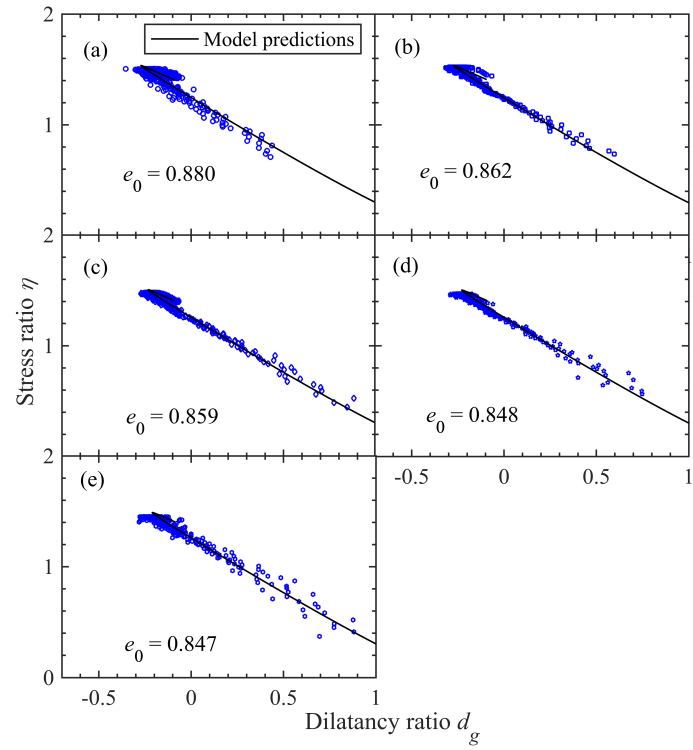

**Fig. 6**. Model simulations of the stress-dilatancy behaviour of loose Karlsruhe fine sand (Wichtmann and Triantafyllidis, 2016)